\documentclass{article}

\usepackage{amssymb,amsfonts,amsmath,stmaryrd}
\usepackage{cite,enumerate,float,indentfirst}
\usepackage{color}

\def\be{\begin{eqnarray}}
\def\ee{\end{eqnarray}}
\def\nn{\nonumber}

\hoffset= - 1.0in         

\begin{document}

\hfill MITP/TH-04/19

\hfill ITEP/TH-04/19

\hfill IITP/TH-04/19

\bigskip

\centerline{\Large{Extension of KNTZ trick to non-rectangular representations
}}

\bigskip

\centerline{{\bf A.Morozov}}

\bigskip

\centerline{\it MIPT, ITEP, \& IITP, Moscow, Russia}

\bigskip

\centerline{ABSTRACT}

\bigskip

{\footnotesize
We claim that the recently discovered universal-matrix precursor
for the $F$ functions, which define the differential expansion
of colored polynomials for twist and double braid knots,
can be extended from rectangular to non-rectangular representations.
This case is far more interesting, because it involves multiplicities
and associated mysterious gauge invariance of arborescent calculus.
In this paper we make the very first step -- reformulate in this form
the previously known formulas for the simplest non-rectangular
representations $[r,1]$  and demonstrate their drastic simplification
after this reformulation.
}

\bigskip

\bigskip

Spectacular success \cite{KNTZ,M19} of the lasting program
\cite{M1605}-\!\!\cite{M333}
to calculate colored knot polynomials \cite{knotpols}
for antiparallel double braids (double twist knots)
and Racah matrices \cite{Racah} in all rectangular representations $R$
from the evolution properties  \cite{DMMSS}-\!\!\cite{Anokhevo}
of their differential expansions \cite{DGR,IMMMfe,evo,diffarth},
opens a way to attack the main problem of arborescent calculus \cite{arbor,arborgauge}:
understanding of non-rectangular representations.
The main difference from rectangular case is that {\it multiplicities} occur
in the product of representations, and this makes the notion of Racah matrices
ambiguous.
In the language of \cite{arborgauge} this is described as the new gauge invariance
and one of the problems is to define gauge-invariant arborescent vertices.
However, before that there is a problem to calculate the Racah matrices
$\bar S$ and $S$, which enter the definition of "fingers" and "propagators",
connected by these vertices.
These problems, are not fully unrelated, because $\bar S$ and $S$
in non-rectangular case are not gauge invariant -- still one can ask what they are
in a particular gauge.
As suggested in \cite{M16}, the key to evaluation of $\bar S$ is
differential expansion (DE) for twist knots \cite{evo} -- which, once known,
straightforwardly produces $\bar S$ for rectangular $R$,
because of spectacular (and still unexplained!)
factorization property of the DE coefficients for double braids.
$S$ are then easily extractable as a diagonalizing matrix for $\bar S$ --
it is enough to solve a system of linear equations.
However, for non-rectangular $R$ the situation is worse:
differential expansion for double braids includes not $\bar S$ itself,
but some gauge-invariant combination of its matrix elements,
and also the linear system for $S$ is degenerate and again provides
only the information about gauge-invariant quantities.
The problem therefore is to extract at this stages exactly the combinations,
needed for arborescent calculus -- and we do not yet know what they are.
In other words, for non-rectangular $R$ we face a whole complex of
related problems, which is partly surveyed in
\cite{arborgauge}, \cite{Mnonrect} and, especially, \cite{MnonrectRacah}.
Whatever  the resolution will be, the first step is going to be
the differential expansion for twist knots --
and it is still not fully known for non-rectangular $R$.
It is the goal of the present paper to suggest a mixture of the results
of \cite{Mnonrect} and \cite{KNTZ,M19} to advance in this direction.

We avoid repeating the whole story and refer to \cite{M19} for the latest summary
and references.
The crucial facts are the observation of \cite{M16} for the antiparallel
double braid in Fig.1:
\be
\boxed{ \begin{array}{ccc}
{\cal H}_{R}^{(m,n)}
= \sum_{X\subset R\otimes\bar R}
Z^X_{R} \cdot {\cal F}_X^{(m,n)}
= \sum_{X\subset R\otimes\bar R}
Z^X_{R} \cdot {F_X^{(m)}F_X^{(n)}}
\\ \\
F_X^{(m)} =     \sum_{Y\triangleleft\, X} f_{XY}\cdot\Lambda_Y^m
\end{array} }
\label{difexpan}
\ee
and the second observation of \cite{KNTZ,M19}, that for rectangular $R$:
\be
\boxed{
F_X^{(m)} = \sum_{Y\triangleleft\, X} \Big({\cal B}^{m+1}\Big)_{XY}
}
\label{FviaB}
\ee
where ${\cal B}$ is a {\it universal} triangular "embedding" matrix
with $Y\triangleleft\, X\subset R\otimes \bar R$.
In this paper we consider the possibility for (\ref{FviaB}) to hold also
for non-rectangular $R$. We do {\it not} discuss what are the $m$-independent
differential combinations $Z^X_R$, which is also a highly non-trivial story
in this case, see \cite{Mnonrect} and a number of preceding papers,
cited therein.
This $Z$-story actually belongs to the theory of
a single figure-eight knot, $4_1$ and is well separated from the problem of
$m$-dependence, which we address now --
though both are equally relevant for the next step towards Racah matrices.

Representations  $X$ and $Y$ in (\ref{difexpan}) are {\it composite}, see Fig.2.
For {rectangular} representations $R=[r^s] = \overbrace{[r,\ldots,r]}^{s \ {\rm times}}$
only very special {\it diagonal} composites $(\lambda,\lambda)$
contribute to $R\otimes\bar R$ -- and they
are in one-to-one correspondence with the Young sub-diagrams of
$\lambda\subset R$, and "embedding" for diagonal composites is understood as embedding
of the corresponding $\lambda$:
\vspace{-0.3cm}
\be
(\mu,\mu)\triangleleft(\lambda,\lambda)\ \Longleftrightarrow\ \mu\subset\lambda
\ee
The entries of the matrix ${\cal B}$ in (\ref{FviaB})
are expressed through the skew Schur functions:
\be
\boxed{
{\cal B}_{\lambda\mu} = (-)^{|\lambda|-|\mu|}\cdot\Lambda_\lambda\cdot
\frac{\chi_\mu^\circ\cdot \chi_{\lambda^\vee/\mu^\vee}^\circ}{\chi_\lambda^{\circ}}
}
\label{Bordinary}
\ee
where  $\vee$ stands for transposition of the Young diagram,
and $\Lambda_\mu$ are the eigenvalues of ${\cal R}$-matrix in the channel
\be
R\otimes\bar R = \oplus_{\mu\in R}\ (\mu,\mu)
\label{embed}
\ee
best expressed through the hook parameters of $\lambda = (a_1,b_1|a_2,b_2|,\ldots)$:
\be
\Lambda_\lambda = \prod_{i}^{\#_{\rm hooks}(\lambda)} (q^{a_i-b_i}A)^{2(a_i+b_i+1)}
\ee
Index $\circ$ means that Schur functions are evaluated at the "unit" locus
in the space of time-variables,
\be
\chi_\lambda^\circ = \chi_\lambda\Big\{p_k = \frac{(q-q)^k}{q^k-q^{-k}}\Big\}
\label{unitlocus}
\ee
At $q=1$ this is equivalent to putting $p_k=\delta_{k,1}$, and there is even a
a special notation for the result: $\chi_\lambda\{\delta_{k,1}\} = d_\lambda$.
The value of skew Schur at the unit locus at $q=1$ can be also expressed through
{\it shifted} Schur functions \cite{Okshift}
\be
\frac{  \chi_{\lambda/\mu}^\circ}{\chi_\lambda^\circ} = \bar\chi_\mu\{\bar p^\lambda\}
\ \ \ \ \Longrightarrow \ \ \ \
{\cal B}_{\lambda \mu } \sim \bar\chi_{\mu }\{\bar p^{ \lambda^\vee }\}
\label{Bviashift}
\ee
evaluated at
\vspace{-0.3cm}
\be
p_k=p_k^\lambda = \sum_{i=1}^{l_\lambda} \Big((\lambda_i-i)^k-(-i)^k\Big)
\label{lamloc}
\ee
where $\lambda_i$ denotes the lengths of $l_\lambda$ lines of the Young diagram $\lambda$.
According to this definition, the shifted $\bar\chi_\mu\{p^\lambda\}$ vanishes
at the $\lambda$-locus (\ref{lamloc}) whenever $\mu$ is {\it not}
a sub-diagram of $\lambda$.
Since shifted Macdonald functions can be defined in just the same way as Schurs \cite{Okmac},
eq.(\ref{Bviashift}) can be immediately used to define a "refined" matrix ${\cal B}$
and thus, through (\ref{FviaB}), the hyper-polynomials
(by definition of \cite{DMMSS} they are result of a {\it clever} substitution of Schur
by Macdonald functions in HOMPLY-PT polynomials, see also \cite{GGS,diffarth} and \cite{NawOb}).
It was demonstrated in \cite{KNTZ} that they are indeed {\it positive} Laurent {\it polynomials},
presumably in all rectangular representations and for all double twist knots.

\bigskip

For non-rectangular $R$ expressions for $Z_X$ and $F_X$ become somewhat complicated,
and one can expect that expression (\ref{FviaB})
of $F_X$ through an auxiliary  matrix ${\cal B}$
once again leads to drastic simplification.
As we will see, {\bf this is indeed the case}.
Note that of the three properties
\be
F_X^{m=-1}=1, \ \ \ \ F_X^{m=0}=0, \ \ \ \
F_X^{m=1}=\prod_{i}^{\#_{\rm hooks}(X)} (-q^{a_i-b_i}A^2)^{ a_i+b_i+1 },
\ee
for the figure-eight knot, unknot and trefoil respectively,
the first one is automatic in (\ref{FviaB}), the second one requires that sum of the entries
of ${\cal B}$ is zero along each line, $\sum_Y{\cal B}_{XY}=0$ $\ \forall X$,
and the third one then says that $\sum_Y({\cal B}^2)_{XY}={\cal B}_{X\emptyset}$
is a monomial $F_X^{(1)}$.

\bigskip

In this paper we consider the simplest case of $R=[r,1]$,
for which the answers are already known from \cite{Mnonrect}.
In this case in addition to the $2r+1$ diagrams $X=(\lambda,\lambda)$ with
$\lambda \subset R=[r,1]$, i.e. $\lambda =\emptyset, [i], [i,1]$, $i=1,\ldots,r$
there are $r-1$ additional composite pairs  $\boxed{\tilde X_i=([i-1,1],[i])\oplus([i],[i-1,1])}$
with the same dimensions and eigenvalues
\vspace{-0.3cm}
\be
\tilde\Lambda_i = (q^{i-2}A)^{2i}, \ \ \ {i=2,\ldots,r}
\ee
each contributing {\it once} to the differential expansion.
These $\tilde X_i$ contribute $r-1$ additional lines to the
matrix ${\cal B}$,
which thus becomes of the size $2r+1+r-1=3r$.
Remarkably, ${\cal B}$ remains {\it triangular},  though a notion of embedding for
generic composites $X$ gets somewhat more subtle than (\ref{embed}).
The first $2r+1$ lines remain as they were in (\ref{Bordinary}).
The new entries in the new $r-1$ lines $\tilde X_i$ with $i=2,\ldots,r$ are:

\newpage
\phantom.
\vspace{-1.0cm}
\be
\boxed{
\begin{array}{ccl}
{\cal B}_{\tilde X_i,\emptyset} = \ \ \ \  &  \frac{(-)^{i+1} \tilde\Lambda_i }{ q^{(i-1)(i-2)}}\cdot A^2
\nn\\ \nn\\
\phantom.\! {\cal B}_{\tilde X_i,[j]} = \ \ \ \
&\frac{(-)^{i+j-1}\tilde\Lambda_i}{q^{(i-1)(i-j)} }\cdot
\frac{[i-2]!}{[i-j]![j-1]!} \cdot
\frac{ [i-1]\cdot q^{3i+j-2} A^2
- [i-j] \cdot q^{i-3} A^2-   [j-1] }{q^{2j}-1} \ \ \ \ & _{j=1,\ldots, i}
\nn\\ \nn\\
{\cal B}_{\tilde X_i,[j,1]} = \ \ \ \  &\frac{(-)^{i+j-1}\tilde\Lambda_i}{q^{i^2-ij-2i-j+7} }
\cdot \frac{[i-2]!}{[i-j-1]![j-1]!} \cdot
\frac{(A^2-q^2)(A^2-q^6)}{(q^{2j+2}-1)(A^2q^{2j-4}-1)} & _{j =1,\ldots, i-1}
\nn\\ \nn\\
{\cal B}_{\tilde X_i,\tilde X_j} = \ \ \ \ & \frac{(-)^{i+j}\tilde\Lambda_i}{q^{(i-1)(i-j)} }\cdot
\frac{[i-2]!}{[i-j]![j-2]!}\cdot\frac{A^2q^{2i-4}-1 }{A^2q^{2j-4}-1}
& _{j =2, \ldots, i} \\
\end{array}
}
\label{newB}
\ee

\bigskip

\noindent
In particular,
\vspace{-0.3cm}
\be
{\cal B}_{\tilde X_i, [1]} = \ \ & (-)^i \tilde\Lambda_i \cdot \frac{[i+1]\cdot A^2  }{q^{(i-2)^2}}
\nn \\ 
{\cal B}_{\tilde X_i,[1,1]} = \ \
&  (-)^i \tilde\Lambda_i\cdot \frac{ A^2 -q^6}{q^{i^2-3i+4}\cdot(q^{4}-1)}
\nn \\ 
{\cal B}_{\tilde X_i,[i]} = \ \ & -\tilde\Lambda_i\cdot \frac{ A^2q^{4i-2}-1   }{q^{2i}-1}
\nn \\ 
{\cal B}_{\tilde X_i,[i,1]} = \ \ & 0
\ee
As we see, these entries essentially depend on $A$, and are therefore sensitive
to characters (or something else) beyond the unit locus (\ref{unitlocus}).

In the simplest case of $R=[2,1]$ the matrix is
\be
{\cal B}^{[2,1]} = \left(\begin{array}{c|cccccc}
& \emptyset & [1] & [1,1] & [2] & [2,1] & \tilde X_{2} \\
&&&&&&\\
\hline
&&&&&&\\
\phantom.\emptyset & 1 & 0 & 0 & 0 & 0 & 0 \\
&&&&&&\\
\phantom.[1] & -A^2 & A^2 & 0 & 0 & 0 & 0\\
&&&&&&\\
\phantom.[1,1] & \frac{A^4}{q^2} & -\frac{[2]A^4}{q^3} & \frac{A^4}{q^4} & 0 & 0 & 0 \\
&&&&&&\\
\phantom.[2] & q^2A^4 & -[2]q^3A^4 & 0 & q^4A^4 & 0 & 0 \\
&&&&&&\\
\phantom.[2,1] & -A^6 & [3]A^6 & -\frac{[3]A^6}{[2]q} & -\frac{[3]qA^6}{[2]} & A^6 & 0 \\
&&&&&&\\
\tilde X_2 & -A^6 & [3]A^6 & \frac{ (A^2-q^6)A^4}{q^2(q^4-1)}
& -\frac{(A^2q^6-1)A^4}{q^4-1}&0&A^4
\end{array}\right)
\ee

\bigskip

\noindent
The new one -- revealed by consideration of the non-rectangular $R$ -- is the last line.\\

For  $R=[3,1]$ the line $\tilde X_2$ remains the same
-- this is the {\it universality} property of ${\cal B}$ --
and there is one more {\it new}, as compared to (\ref{Bordinary}), line for $\tilde X_3$:

\bigskip

\centerline{{\footnotesize
$
\begin{array}{c||ccccccccc}
&\emptyset&[1]&[1,1]&[2]&[2,1]&[3]&[3,1]&\tilde X_2&\tilde X_3 \\
&&&&&&&&&\\
\hline
&&&&&&&&&\\
\tilde X_3 & q^4A^8 & -[4]q^5A^8 & -\frac{q^2(A^2-q^6)A^6}{q^4-1} &
\frac{q^4\big(A^2(q^{10}+q^8-1)-1\big)A^6}{q^4-1} & \frac{q^4(A^2-q^2)(A^2-q^6)A^6}{(q^6-1)(A^2-1)}
& -\frac{q^6(A^2q^{10}-1)A^6}{q^6-1} & 0 & -\frac{q^4(A^2q^2-1)A^6}{A^2-1}& q^6A^6
\end{array}
$
}}

\bigskip

\noindent
One can compare with the original formulas for $F_X$ in \cite{Mnonrect}
to appreciate the simplification.

To make the story about $R=[r,1]$ complete, we need also explicit formula
for the $Z$-factors.
They are made from the {\it differentials} $D_n:=Aq^N-A^{-1}q^{-N}$,
for example for $R=[2,1]$
\be
Z^\emptyset_{[2,1]}=1, \ \ \ Z^{[1]}_{[2,1]} = \frac{[3]D_0^2+[3]^2D_2D_{-2}}{[2]^2}, \  \ \
Z^{[2]}_{[2,1]} = \frac{[3]}{[2]}\cdot D_3D_2D_0D_{-2}, \ \ \
Z^{[1,1]}_{[2,1]}=\frac{[3]}{[2]}\cdot D_2D_0D_{-2}D_{-3}, \nn \\
Z_{[2,1]}^{[2,1]} = D_3D_2D_1D_{-1}D_{-2}D_{-3}, \ \ \ Z^{\widetilde{X_2}}_{[2,1]} =
-[3]^2(q-q^{-1})^4\cdot D_2D_{-2} \ \ \ \ \
\ee
We see that one of the $Z$-factors is not fully factorized -- this is the one,
associated with the diagram $\lambda=[1]$ which has non-trivial multiplicity.
Since multiplicity is two, it is naturally decomposed into sum of two factorized items.
For a full understanding of how this works we need a more sophisticated theory,
involving the analogue of $U$-matrices from \cite{M19} for non-rectangular $R=[2,1]$
and reduction from the full-fledged $10\times 10$ matrix representation to the $6\times 6$ one.
This is a difficult and still not fully-developed subject beyond the scope of the
present paper.

Known at this stage are all the $Z$-factors in the case of $R=[r,1]$, see \cite{Mnonrect}.
They are not-quite-factorized for all single-line $\lambda=[k]$ with, which enter with
multiplicities two:
\be
Z_{[r,1]}^{[k]} =\frac{[r+1]!}{[r][k]![r-k]!}\cdot \frac{D_{r+k-2}!D_{k-2}!}{D_{r-1}!}
\left(D_{r-2}D_{k-1}-[r+1][k](q-q^{-1})^2\right)
= \nn \\
=  \frac{[r+1]}{[r]^2}\left(\frac{[r]!}{[k+1]![r-k-1]!}\cdot
\frac{D_{r+k-2}!D_{k-1}!}{D_{r-1}!D_{-1}!}\cdot D_{r-2} +
[k]\cdot \frac{[r+1]!}{[k+1]![r-k]! } \cdot
\frac{D_{r+k-1}!D_{k-2}!}{D_{r-1}!D_{-1}!}\cdot D_{-2}\right)
\label{Zkr1}
\ee
and are factorized for the other two series:
\be
Z_{[r,1]}^{[k,1]} = \frac{[k]\,[r+1]!}{[r]\,[k+1]![r-k]!}\cdot
\frac{D_{r+k-1}!D_{k-1}!}{D_{r-1}!D_{k-2}}
D_{-1}D_{-2}D_{-3}
\ee
\be
Z_{[r,1]}^{\widetilde{X_k}} =  -(q-q^{-1})^4\cdot\frac{[r+1]^2\,[r-1]!}{ [r-k]![k-2]!}
\cdot \frac{D_{r+k-2}!D_{k-3}!}{D_{r-1}!}\cdot D_{-2}
\ee
In fact, (\ref{Zkr1}) is also factorized (the first term in the second line vanishes)
at $k=r$.

\bigskip

For generic $R$ we should consider all
composite $\boxed{X=\oplus_{|\lambda|=|\lambda'|}(\lambda,\lambda')}$
with all pairs of the same-size sub-diagrams of $R$:  $\lambda,\lambda'\subset R$,
$|\lambda|=|\lambda'|$.
For example, for $R=[3,2]$ there will be three non-diagonal ($\lambda'\neq\lambda$) pairs:
the two already familiar $([2],[1,1])\oplus([1,1],[2])$, $([3],[2,1])\oplus([2,1],[3])$
and a new one: $([3,1],[2,2])\oplus([2,2],[3,1])$.
For the psychologically important $R=[4,2]$
we encounter the first triple $\{[4],[3,1],[2,2]\}$, giving rise to three pairs
$([4],[3,1])\oplus([3,1],[4])$,
 $([3,1],[2,2])\oplus([2,2],[3,1])$ and $([4],[2,2])\oplus([2,2],[4])$.
Formulas (\ref{newB}) should be straightforwardly extendable to this general case --
but it remains to be done, and it remains to be seen if {\it triangular} shape of ${\cal B}$
persists.
Hopefully straightforward are also their Macdonald deformations -- and it is interesting
to see if this leads to hyper-{\it polynomials}, but {\it not} fully positive --
as currently suspected for non-rectangular representations.

\section*{Acknowledgements}

My work is partly supported by the grant of the
Foundation for the Advancement of Theoretical Physics BASIS,
by RFBR grant 19-02-00815
and by the joint grants 17-51-50051-YaF, 18-51-05015-Arm,
18-51-45010-Ind,  RFBR-GFEN 19-51-53014.

\newpage

\bigskip

\begin{figure} \label{doublebraid}
\begin{picture}(200,350)(-220,-300)
\qbezier(-40,0)(-50,20)(-60,0)
\qbezier(-40,0)(-50,-20)(-60,0)
\qbezier(-20,0)(-30,20)(-40,0)
\qbezier(-20,0)(-30,-20)(-40,0)
\qbezier(-20,0)(-15,10)(-10,10)
\qbezier(-20,0)(-15,-10)(-10,-10)
\put(-5,0){\mbox{$\ldots$}}
\qbezier(10,10)(15,10)(20,0)
\qbezier(10,-10)(15,-10)(20,0)
\qbezier(20,0)(30,20)(40,0)
\qbezier(20,0)(30,-20)(40,0)
\qbezier(40,0)(50,20)(60,0)
\qbezier(40,0)(50,-20)(60,0)
\put(-60,0){\line(-1,2){10}}
\put(-60,0){\line(-1,-2){10}}
\put(60,0){\line(1,2){10}}
\put(60,0){\line(1,-2){10}}
\qbezier(0,-80)(-20,-90)(0,-100)
\qbezier(0,-80)(20,-90)(0,-100)
\qbezier(0,-100)(-20,-110)(0,-120)
\qbezier(0,-100)(20,-110)(0,-120)
\qbezier(0,-120)(-10,-125)(-10,-130)
\qbezier(0,-120)(10,-125)(10,-130)
\put(0,-145){\mbox{$\vdots$}}
\qbezier(0,-160)(-10,-155)(-10,-150)
\qbezier(0,-160)(10,-155)(10,-150)
\qbezier(0,-160)(-20,-170)(0,-180)
\qbezier(0,-160)(20,-170)(0,-180)
\qbezier(0,-180)(-20,-190)(0,-200)
\qbezier(0,-180)(20,-190)(0,-200)
\put(0,-80){\line(-2,1){10}}
\put(0,-80){\line(2,1){10}}
\put(0,-200){\line(-2,-1){10}}
\put(0,-200){\line(2,-1){10}}
\put(0,-200){\line(-2,-1){20}}
\put(0,-200){\line(2,-1){20}}
\qbezier(-10,-75)(-80,-40)(-70,-20)
\qbezier(10,-75)(80,-40)(70,-20)
\put(-10,-205){\vector(2,1){2}}
\put(10,-205){\vector(2,-1){2}}
\put(-65,10){\vector(-1,2){2}}
\put(65,10){\vector(-1,-2){2}}
\put(-70,-20){\vector(1,2){2}}
\put(70,-20){\vector(1,-2){2}}
\put(-3,20){\mbox{\footnotesize$2n$}}
\put(-32,-140){\mbox{\footnotesize $2m$}}
%
\qbezier(-70,20)(-80,40)(-97,25)
\qbezier(-97,25)(-111,13))(-100,-30)
\qbezier(-100,-30)(-60,-230)(-20,-210)
\qbezier(70,20)(80,40)(97,25)
\qbezier(97,25)(111,13))(100,-30)
\qbezier(100,-30)(60,-230)(20,-210)
\put(-102,-22){\vector(1,-4){2}}
\put(100,-30){\vector(1,4){2}}

\put(-200,-270){\mbox{
$
\boxed{
\begin{array}{ccc}
{\cal H}_{R}^{(m,n)}
=   \sum_{Y,Y'\subset R\otimes R}
\frac{\sqrt{{\cal D}_{Y}{\cal D}_{Y'}}}{D_R(N)}
\,\bar S_{YY'}^{R}\, \Lambda_Y^m\Lambda_{Y'}^n
= \sum_{X\subset R\otimes\bar R}
Z^X_{R} \cdot {\cal F}_X^{(m,n)}
= \sum_{X\subset R\otimes\bar R}
Z^X_{R} \cdot {F_X^{(m)}F_X^{(n)}}
 \\ \\
 F_X^{(m)} =     \sum_{Y\triangleleft\, X} f_{XY}\cdot\Lambda_Y^m
\end{array}
}
$
}}

\end{picture}

\caption{Antiparallel double braid and the two representations of associated
HOMFLY-PT polynomial: arborescent one through exclusive Racah matrix $\bar S$
from \cite{arbor}
and the factorized differential expansion from \cite{M16}.
}
\end{figure}
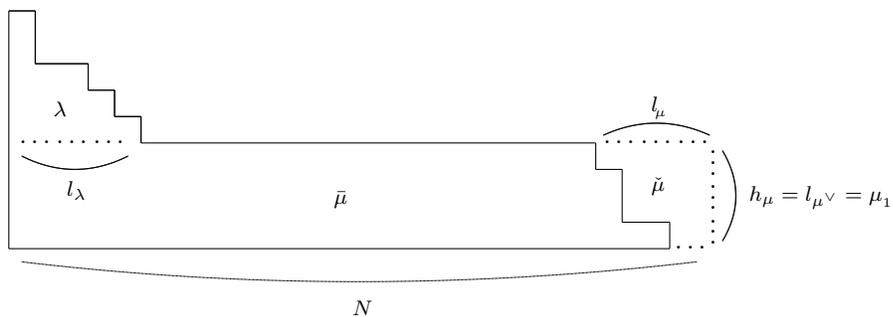
\begin{figure} \label{comporep}

\begin{picture}(300,30)(-90,-30)

\put(0,0){\line(0,1){90}}
\put(0,0){\line(1,0){250}}
\put(50,40){\line(1,0){172}}

\put(0,90){\line(1,0){10}}
\put(10,90){\line(0,-1){20}}
\put(10,70){\line(1,0){20}}
\put(30,70){\line(0,-1){10}}
\put(30,60){\line(1,0){10}}
\put(40,60){\line(0,-1){10}}
\put(40,50){\line(1,0){10}}
\put(50,50){\line(0,-1){10}}

\put(265,2){\mbox{$\vdots$}}
\put(265,15){\mbox{$\vdots$}}
\put(265,28){\mbox{$\vdots$}}

\put(252,0){\mbox{$\ldots$}}
\put(253,40){\mbox{$\ldots$}}
\put(239,40){\mbox{$\ldots$}}
\put(225,40){\mbox{$\ldots$}}

\put(222,40){\line(0,-1){10}}
\put(222,30){\line(1,0){10}}
\put(232,30){\line(0,-1){20}}
\put(232,10){\line(1,0){18}}
\put(250,0){\line(0,1){10}}

\put(0,90){\line(1,0){10}}
\put(10,90){\line(0,-1){20}}
\put(10,70){\line(1,0){20}}
\put(30,70){\line(0,-1){10}}
\put(30,60){\line(1,0){10}}
\put(40,60){\line(0,-1){10}}
\put(40,50){\line(1,0){10}}
\put(50,50){\line(0,-1){10}}


{\footnotesize
\put(123,17){\mbox{$ \bar \mu$}}
\put(17,50){\mbox{$\lambda$}}
\put(243,22){\mbox{$\check \mu$}}
\qbezier(270,3)(280,20)(270,37)
\put(280,18){\mbox{$h_\mu = l_{\mu^{\vee}}=\mu_{_1}$}}
\qbezier(5,-5)(132,-20)(260,-5)
\put(130,-25){\mbox{$N $}}
\qbezier(5,35)(25,25)(45,35)
\put(22,20){\mbox{$l_\lambda$}}
\qbezier(225,43)(245,52)(265,43)
\put(243,52){\mbox{$l_{\!_\mu}$}}
}

\put(4,40){\mbox{$\ldots$}}
\put(18,40){\mbox{$\ldots$}}
\put(32,40){\mbox{$\ldots$}}

\put(0,-80){\mbox{
$(\lambda,\mu)= \Big[\lambda_1+\mu_1,\ldots,\lambda_{l_\lambda}+\mu_1,\
\underbrace{\mu_1,\ldots,\mu_1}_{N-l_{\!_\lambda}-l_{\!_\mu}},\
\mu_1-\mu_{_{l_{\!_\mu}}},\ldots,\mu_1-\mu_2\Big]$
}}

\end{picture}

\caption{Composite representation of $Sl_N$, described by the $N$-dependent
Young diagram
  }

\end{figure}


\begin{thebibliography}{12}




\bibitem{KNTZ}
M.Kameyama, S.Nawata, R.Tao, H.D.Zhang,  arXiv:1902.02275

\bibitem{M19} A.Morozov, Phys.Lett.  B 793 (2019) 116-125,  arXiv:1902.04140



\bibitem{M1605} A.Morozov,  Nucl.Phys. B 911 (2016) 582-605, arXiv:1605.09728

\bibitem{M16} A.Morozov,  JHEP 1609 (2016) 135, arXiv:1606.06015 v8




\bibitem{KM17fe}  Ya.Kononov and A.Morozov,
Theor.Math.Phys. 193 (2017) 1630-1646,
arXiv:1609.00143

\bibitem{KM17tw} Ya.Kononov and A.Morozov,
Mod.Phys.Lett. A Vol. 31, No. 38 (2016) 1650223,
arXiv:1610.04778




\bibitem{Mnonrect} A.Morozov, Mod.Phys.Lett. A33 No. 12 (2018) 1850062, arXiv:1612.00422

\bibitem{MnonrectRacah}
A.Morozov, Phys.Lett. B 766 (2017) 291-300, arXiv:1701.00359

\bibitem{M333} A.Morozov,  Phys.Lett. B778 (2018) 426-434,  arXiv:1711.09277


\bibitem{knotpols}
J.W.Alexander, Trans.Amer.Math.Soc. 30 (2) (1928) 275-306\\
V.F.R.Jones, Invent.Math. 72 (1983) 1 Bull.AMS 12 (1985) 103 Ann.Math. 126 (1987) 335\\
L.Kauffman, Topology 26 (1987) 395\\
P.Freyd, D.Yetter, J.Hoste, W.B.R.Lickorish, K.Millet, A.Ocneanu, Bull. AMS. 12 (1985) 239\\
J.H.Przytycki and K.P.Traczyk, Kobe J Math. 4 (1987) 115-139\\
A.Morozov, Theor.Math.Phys. 187 (2016) 447-454, arXiv:1509.04928

\bibitem{Racah}
G. Racah, 
Phys.Rev. {\bf 62} (1942) 438-462\\
E.P. Wigner, Manuscript, 1940,  in: {\sl Quantum Theory of Angular Momentum},
pp. 87–133, Acad.Press, 1965;
{\sl Group Theory and Its Application to the Quantum Mechanics of Atomic Spectra},
Acad.Press,   1959\\
L.D. Landau and E.M. Lifshitz, {\sl Quantum Mechanics: Non-Relativistic Theory},
Pergamon Press, 1977 \\
J. Scott Carter, D.E. Flath, M. Saito, {\sl The Classical and Quantum 6j-symbols},
Princeton Univ.Press, 1995 \\
S. Nawata, P. Ramadevi and Zodinmawia, Lett.Math.Phys. {\bf 103} (2013) 1389-1398,
arXiv:1302.5143 \\
A. Mironov, A. Morozov, A. Sleptsov,  	JHEP 07 (2015) 069, arXiv:1412.8432


\bibitem{DMMSS}
M.Aganagic, Sh.Shakirov, arXiv:1105.5117 \!\!;  arXiv:1202.2489 \!\!; arXiv:1210.2733 \\
P.Dunin-Barkowski, A.Mironov, A.Morozov, A.Sleptsov, A.Smirnov, JHEP 03 (2013) 021,
arXiv:1106.4305 \\
I. Cherednik, arXiv:1111.6195

\bibitem{IMMMfe} H. Itoyama, A. Mironov, A. Morozov and An. Morozov, JHEP 2012 (2012) 131,
arXiv:1203.5978

\bibitem{evo} A. Mironov, A. Morozov and An. Morozov,
AIP Conf. Proc. 1562 (2013) 123, arXiv:1306.3197 \!\!;
Mod. Phys. Lett. A 29 (2014) 1450183,  arXiv:1408.3076

\bibitem{pretzel}
D.Galakhov, D.Melnikov, A.Mironov, A.Morozov, A.Sleptsov,
 Phys.Lett. B743 (2015) 71, arXiv:1412.2616 \\
A.Mironov, A.Morozov, A.Sleptsov,
JHEP 07 (2015) 069, arXiv:1412.8432 \\
D.Galakhov, D.Melnikov, A.Mironov and A.Morozov,
Nucl.Phys. B 899 (2015) 194-228, arXiv:1502.02621 \\
A.Mironov, A.Morozov, An.Morozov, A.Sleptsov,
JETP Lett. 104 (2016) 56-61, Pisma Zh.Eksp.Teor.Fiz. 104 (2016) 52-57,  arXiv:1605.03098 \\
Sh. Shakirov and A. Sleptsov,
arXiv:1611.03797 \\
S.Arthamonov and Sh.Shakirov,  arXiv:1704.02947

\bibitem{Anokhevo}
A.Anokhina, A.Morozov, JHEP 1804 (2018) 066, arXiv:1802.09383  \\
P.Dunin-Barkowski, A.Popolitov, S.Popolitova, arXiv:1812.00858 \\
A.Anokhina, A.Morozov, A.Popolitov,  arXiv:1904.10277

\bibitem{DGR}
N.M.Dunfield, S.Gukov  and J.Rasmussen, math/0505662


\bibitem{diffarth} S.Arthamonov, A.Mironov, A.Morozov,
Theor.Math.Phys. 179 (2014) 509-542, arXiv:1306.5682



\bibitem{arbor}
A.Mironov, A.Morozov, An.Morozov, P.Ramadevi, V.K.Singh,
JHEP {\bf 1507} (2015) 109,  arXiv:1504.00371 \\
S.Nawata, P.Ramadevi, V.K.Singh,  arXiv:1504.00364 \\
A.Mironov and A.Morozov, Phys.Lett. B755 (2016) 47-57, arXiv:1511.09077

\bibitem{arborgauge}
A. Mironov, A. Morozov, An. Morozov, P. Ramadevi, V.K. Singh and A. Sleptsov,
J.Phys. A: Math.Theor. {\bf 50} (2017) 085201, arXiv:1601.04199

\bibitem{Okshift}
A.Okounkov, G.Olshanksy, Algebra i Analiz 9 (1997) No.2;
Math.Res.Lett. 4 (1997) 69-78, q-alg/9608020

\bibitem{Okmac} A.Okounkov, arXiv:q-alg/9608021

\bibitem{GGS} E.Gorsky, S.Gukov,  M.Stosic,
Fundamenta Mathematicae 243 (2018) 209–299, arXiv:1304.3481

\bibitem{NawOb} S.Nawata and A.Oblomkov,
Contemp. Math. 680 (2016) 137, arXiv:1510.01795




\end{thebibliography}
\end{document}